\documentclass[fleqn,10pt]{wlscirep}
\title{Quantum chemistry and charge transport in biomolecules with superconducting circuits}

\author[1,*]{L. Garc\'ia-\'Alvarez}
\author[1]{U. Las Heras}
\author[1,2]{A. Mezzacapo}
\author[1]{M. Sanz}
\author[1,3]{E. Solano}
\author[1]{L.~Lamata}
\affil[1]{Department of Physical Chemistry, University of the Basque Country UPV/EHU, Apartado 644, E-48080 Bilbao, Spain}
\affil[2]{IBM T. J. Watson Research Center, Yorktown Heights, NY 10598, USA}
\affil[3]{IKERBASQUE, Basque Foundation for Science, Maria Diaz de Haro 3, 48013 Bilbao, Spain}

\affil[*]{garcia.alvarez.la@gmail.com}

\begin{abstract}
We propose an efficient protocol for digital quantum simulation of quantum chemistry problems and enhanced digital-analog quantum simulation of transport phenomena in biomolecules with superconducting circuits. Along these lines, we optimally digitize fermionic models of molecular structure with single-qubit and two-qubit gates, by means of Trotter-Suzuki decomposition and Jordan-Wigner transformation. Furthermore, we address the modelling of system-environment interactions of biomolecules involving bosonic degrees of freedom with a digital-analog approach. Finally, we consider gate-truncated quantum algorithms to allow the study of environmental effects.
\end{abstract}

\begin{document}

\flushbottom
\maketitle
\thispagestyle{empty}

The field of quantum chemistry arises from the application of quantum mechanics in physical models to explain the properties of chemical and biological systems~\cite{Szabo96,Helgaker00}. The study of complex electronic structures in atoms and molecules encounters the difficulty of the exponential growth of the Hilbert space dimensions with the system size~\cite{Sherrill10,Whitfield13}. This fact limits the results reachable with current computers and classical algorithms, and strongly suggests to explore the possibilities of new quantum-based tools~\cite{Aspuru5,Kassal11}.

Quantum simulations are a powerful field based on the imitation of the dynamics of a quantum system in a controllable quantum platform~\cite{Feynman82,Lloyd96}. Theoretical and experimental efforts for solving problems in physical chemistry have been performed in technologies such as NMR~\cite{Du10}, trapped ions~\cite{Lamata14,Casanova14,Shen15}, photonic systems~\cite{Lanyon10,Peruzzo14,Peropadre14}, and superconducting circuits~\cite{Mostame12}, among others. Quantum algorithms for the simulation of electronic structures with fermionic degrees of freedom and its optimisation have been widely studied~\cite{Aspuru08,Whitfield11,Toloui13,Hastings14,Babbush14,Poulin14,Whitfield15,Babbush15}. Environmental effects also play a crucial role in quantum physics, chemistry and biology~\cite{QAspectsLife,ChargeEnergyTrans}. Fundamental phenomena such as electronic transport and electron transfer are described through the correlated dynamics of electrons and phonons, involving bosonic and fermionic modes.

Circuit quantum electrodynamics (cQED) is a cutting-edge technology in terms of design versatility, coherent control, and scalability~\cite{Devoret13}. Indeed, remarkable experimental progress in cQED has enabled the realisation of digital quantum simulations of fermions \cite{Barends15}, spin systems \cite{Salathe15}, and adiabatic quantum computing \cite{Barends15bis}. These aspects, along with the possibility of encoding both fermions and bosons in this platform via digital~\cite{Urtzi14,Mezzacapo14,Urtzi15,Chiesa15} and digital-analog techniques~\cite{Laura15}, make cQED a suitable platform for simulating electronic Hamiltonians~\cite{OMalley15} and dissipative processes.

In this manuscript, we combine efficient digital quantum simulation techniques for electronic Hamiltonians with existing algorithms in quantum chemistry, and we analyze the scalability and feasibility according to the state-of-the-art cQED~\cite{Devoret13}. In this sense, we study the gate fidelities required for the proposed tasks and the error propagation. We extend these procedures by exploiting the possibility of mimicking bosons in superconducting circuits taking full advantage of the multimode spectrum of superconducting transmission lines~\cite{Egger13,Krimer14,Nigg12,Sundaresan15,McKay15}, and propose digital-analog quantum simulations of electron transfer and electronic transport in biomolecules~\cite{Bulla06,Tornow06,Cuniberti09}.

\section*{Results}
\subsection*{Simulation of electronic Hamiltonians}
The electronic structure is a quantum chemistry many-body problem that is usually difficult to solve due to the exponential growth of the Hilbert space with the size of the system. Typically, the aim is to compute ground-state energies and their associated eigenvectors of these interacting electron systems in a fixed nuclear potential.

Among the variety of methods for simulating fermionic models with quantum technologies, one of the most studied approaches considers quantum algorithms using the second quantized formalism of electronic systems~\cite{Aspuru08,Whitfield11,Hastings14,Babbush14,Poulin14,Whitfield15,Babbush15}. The associated Hamiltonian may be represented in different bases, leading to different methods of encoding and scaling improvements in the number of qubits and gates required~\cite{Hastings14,Babbush15}. Furthermore, other approaches related to the Configuration Interaction (CI) matrix have been recently studied~\cite{Toloui13}.

\begin{figure}[t!!!]
\centering
\includegraphics[width=18.2cm]{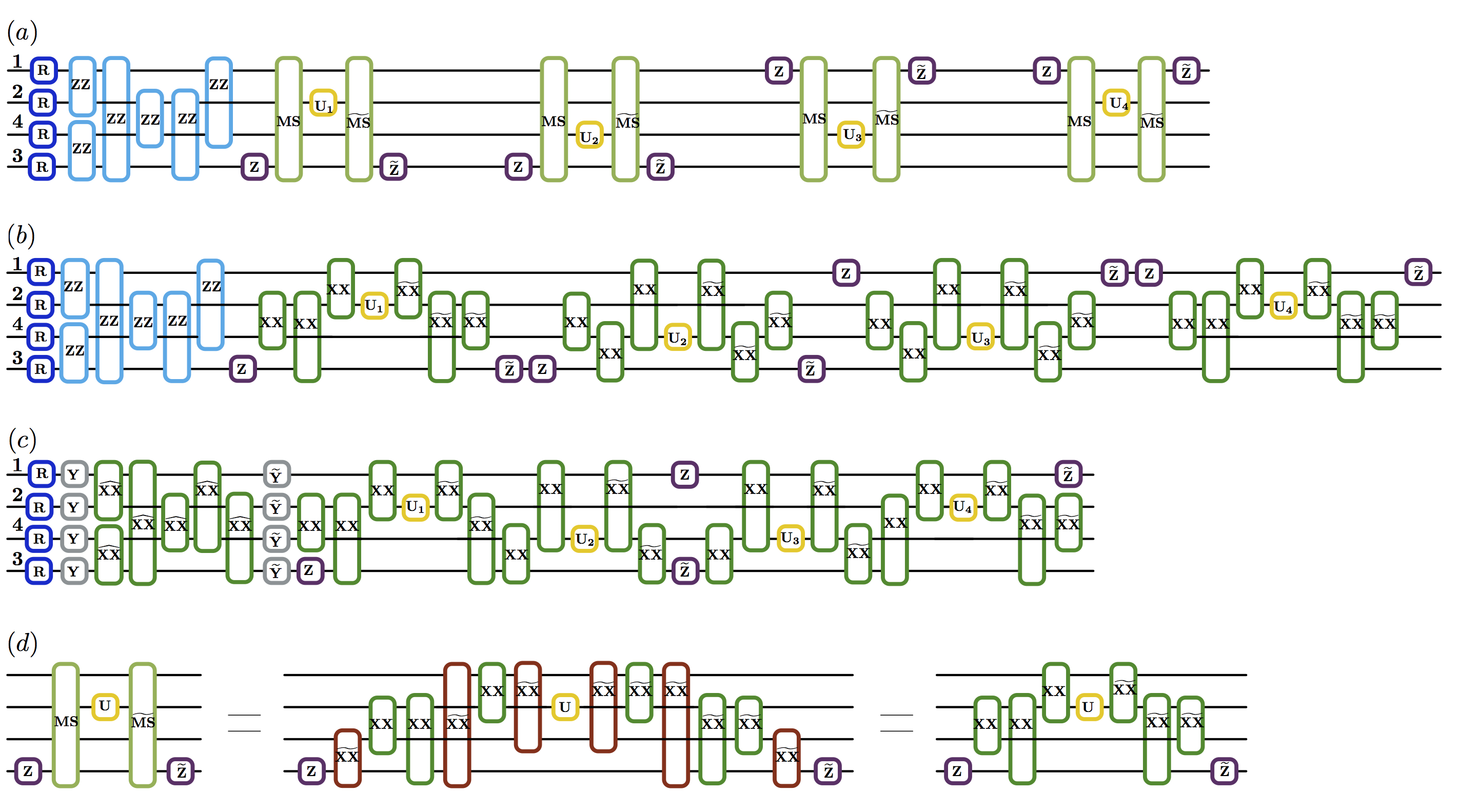}
\caption{Sequence of gates in a single Trotter step of the digital quantum simulation of the Hamiltonian in equation~(\ref{HamSpin}) describing the ${\rm H}_2$ molecule. Notice that, for the sake of optimising the number of gates, we swap the logic label of the third and fourth physical qubits. (a) Original protocol including M\o lmer-S\o rensen~(MS) multiqubit gates (light green), $MS=\exp(i\frac{\pi}{8}S_x^2)$, $\widetilde{MS}=\exp(-i\frac{\pi}{8}S_x^2)$, where $S_x=\sum_i \sigma^x_i$, and intermediate unitaries (yellow) $U_1=U_3=\exp(-i 2h_D t \sigma_j^z)$, $U_2=U_4=\exp(i 2h_D t \sigma_j^z)$, where the subindex $j$ means that it is applied to the $j$-th qubit. The Trotter step also contains $Z_{\pi/2}$-rotations (purple) $Z=\exp(-i\frac{\pi}{4}\sigma_j^z)$ and $\tilde{Z}=\exp(i\frac{\pi}{4}\sigma_j^z)$, single-qubit gates (dark blue) $R= \exp(-i\phi_{j} t \sigma^z_{j})$, with $\phi_1= 4h_{11}+2h_A+4h_C-h_D$, $\phi_2= 4h_{22}+2h_A+4h_C-h_D$, $\phi_3=4h_{33}+2h_B+4h_C-h_D$, and $\phi_4=4h_{44}+2h_B+4h_C-h_D$, and two-qubit gates (light blue) $ZZ=\exp(-i\theta_{ij} t \sigma_i^z \sigma_j^z)$, with the phase $\theta_{ij}$ depending on the qubits involved, such that $\theta_{12}=2h_A$, $\theta_{13}=\theta_{24}=2h_C-h_D$, $\theta_{14}=\theta_{23} =2h_C$, and $\theta_{34}=2h_B$. (b) Trotter step with MS multiqubit gates decomposed into two-qubit gates (dark green). Here, $XX=\exp(-i\frac{\pi}{4}\sigma^x_i \sigma^x_j)$, and $\widetilde{XX}=\exp(i\frac{\pi}{4}\sigma_i^x \sigma_j^x)$.  (c) Optimised Trotter step, in which we have expressed the algorithm in terms of $XX$ gates via the $Y_{\pi/2}$-rotations (grey) $Y=\exp(-i\frac{\pi}{4}\sigma_j^y)$ and $\tilde{Y}=\exp(i\frac{\pi}{4}\sigma_j^y)$, and with $\widehat{XX}=\exp(-i\theta_{ij} t \sigma_i^x \sigma_j^x)$. (d) After a complete decomposition into two-qubit gates of the M\o lmer-S\o rensen gates involved in the simulation of a multiqubit interaction, a simplification of two-qubit gates (red) cancelling each other is shown.}
\label{Tstep}
\end{figure}

The generic Hamiltonian describing a molecular electronic structure consists of the electron kinetic energy term, two-electron Coulomb interactions, and the electron-nuclei potential energy representing the electronic environment~\cite{Szabo96,Helgaker00}. This Hamiltonian in second quantization may be written as
\begin{equation}
H_{\rm e} = \sum_{i,j} h_{ij} c^\dagger_ic_j+\frac{1}{2}\sum_{i,j,k,l}h_{ijkl}c^\dagger_ic^\dagger_jc_kc_l ,
\label{electronHam}
\end{equation}
where the operators $c_i^{\dag}$ and $c_i$ stand for the electrons and obey the fermionic anticommutation relations. Coefficients $h_{ij}$ come from the single-electron integrals of the electron kinetic terms and electron-nuclei interactions, and $h_{ijkl}$ correspond to the two-electron integrals associated with the electron-electron Coulomb interaction. That is, it is expressed in atomic units as
\begin{eqnarray}
h_{ij} \equiv && \int d{\bf r} \varphi^{*}_i ({\bf r})\left( -\frac{1}{2}\nabla^2_{r} - \sum_{k} \frac{Z_k}{|r - R_k |}\right) \varphi_j ({\bf r})  , \\ 
h_{ijkl} \equiv && \int d{\bf r}_1 d{\bf r}_2 \frac{\varphi^{*}_i ({\bf r}_1) \varphi^{*}_j ({\bf r}_2) \varphi_k ({\bf r}_2) \varphi_l ({\bf r}_1)}{| r_1 - r_2 |} ,
\end{eqnarray}
where $R_k$ are nuclear coordinates, $r$ electronic coordinates, and $Z$ the atomic number representing the nuclear charge. Here, $\{\varphi_i ({\bf r})\}$ defines a set of spin orbitals, and ${\bf r}=(r,\sigma)$ denotes the pair of spatial and spin parameters.

Optimal strategies of computation for quantum chemistry merge quantum simulation and classical numerical techniques. These methods, that we name as algorithmic quantum simulation~\cite{Unai16}, allow us to employ quantum simulators for the computationally hard tasks, such as time evolution, on top of the classical algorithm, which provides flexibility for computing relevant observables. In the context of quantum chemistry, we have the example of ground state finding via a variational eigensolver~\cite{Peruzzo14,Casanova14,Wecker15,Bauer15,McClean16}.

The simulation of the dynamics associated with the electronic Hamiltonian in equation~(\ref{electronHam}) involves fermionic operators. Computations with fermionic degrees of freedom in superconducting circuits require the encoding of fermionic operators and their anticommutative algebra in the natural variables of this quantum platform. The Jordan-Wigner transformation~\cite{JordanWigner} maps the fermionic operators into spin-$1/2$ operators, which gives us the qubit representation of the Hamiltonian. In the case of a hydrogen molecule, considering four electronic orbitals, the relations can be written as
\begin{eqnarray}
\label{JWtrans}
c^{\dag}_1 &=& \sigma^{+}_1\mathbb{I}_2\mathbb{I}_3\mathbb{I}_4 , \nonumber \\
c^{\dag}_2 &=& \sigma^{z}_1\sigma^{+}_2\mathbb{I}_3\mathbb{I}_4 , \nonumber \\
c^{\dag}_3 &=& \sigma^{z}_1\sigma^{z}_2\sigma^{+}_3\mathbb{I}_4 , \nonumber \\
c^{\dag}_4 &=& \sigma^{z}_1\sigma^{z}_2\sigma^{z}_3\sigma^{+}_4 . \nonumber  \\
\end{eqnarray}

After this mapping, the Hamiltonian of equation~(\ref{electronHam}) for the ${\rm H}_2$ molecule is rewritten in terms of spin-$1/2$ operators considering only the nonzero coefficients $h_{ij}$ and $h_{ijkl}$, which are computed classically with polynomial resources~\cite{Whitfield11},
\begin{eqnarray}
\label{HamSpin}
H=&&\frac{1}{8}\big[ (4h_{11}+2h_A+4h_C-h_D)\sigma^z_1+\nonumber\\
&&(4h_{22}+2h_A+4h_C-h_D)\sigma^z_2+\nonumber\\
&&(4h_{33}+2h_B+4h_C-h_D)\sigma^z_3+\nonumber\\
&&(4h_{44}+2h_B+4h_C-h_D)\sigma^z_4+\nonumber\\
&&2h_A\sigma^z_1\sigma^z_2+(2h_C-h_D)\sigma^z_1\sigma^z_3+2h_C\sigma^z_1\sigma^z_4+\nonumber\\
&&2h_C\sigma^z_2\sigma^z_3+(2h_C-h_D)\sigma^z_2\sigma^z_4+2h_B\sigma^z_3\sigma^z_4+\nonumber\\
&&2h_D(\sigma^x_1\sigma^y_2\sigma^y_3\sigma^x_4+\sigma^y_1\sigma^x_2\sigma^x_3\sigma^y_4-\nonumber\\
&&\sigma^x_1\sigma^x_2\sigma^y_3\sigma^y_4-\sigma^y_1\sigma^y_2\sigma^x_3\sigma^x_4)\big],
\end{eqnarray}
where
\begin{eqnarray}
h_A&&=h_{1221}=h_{2112},\nonumber\\
h_B&&=h_{3443}=h_{4334},\nonumber\\
h_C&&=h_{1331}=h_{3113}=h_{1441}=h_{4114}=h_{2332}\nonumber\\
&&=h_{3223}=h_{2442}=h_{4224},\nonumber\\
h_D&&=h_{1243}=h_{2134}=h_{1423}=h_{4132}=h_{2314}\nonumber\\
&&=h_{3241}=h_{3421}=h_{4312}=h_{1313}=h_{2424}.
\end{eqnarray}

In general, an analog quantum simulation of an arbitrary Hamiltonian evolution is a difficult problem~\cite{Tian13,Tian15}, since one cannot straightforwardly map the dynamics of a given simulated system onto a given quantum platform. The flexibility and universality of digital quantum simulations allows us to reproduce models that do not appear naturally in a quantum platform. This is done via an expansion of the quantum evolution into discrete steps of quantum gates~\cite{Suzuki90}. An additional advantage of such digital quantum simulations, in the spirit of gate-based quantum algorithms, is their possible improvement with quantum error correction techniques~\cite{Trout15,Barends2014}.

We consider the digital quantum simulation of the ${\rm H}_2$ molecule via the Trotter expansion, which consists in dividing the evolution time $t$ into $l$ time intervals of length $t/l$, and applying sequentially the evolution operator of each term of the Hamiltonian for each time interval~\cite{Lloyd96,Suzuki90,Urtzi15}. The expression of this expansion for a Hamiltonian of the form $H = \sum_j H_j$ reads
\begin{equation}
e^{-iHt} \approx \bigg(\prod_j e^{-iH_j t/l}\bigg)^l ,
\end{equation}
for large $l$, where the dominating error component is $\sum_{i>j} \left[H_i,H_j\right] t^2/2l$, which depends on the value of the commutators and scales with $t^2/l$.

\begin{figure}[t!!!]
\centering
\includegraphics[width=9.5cm]{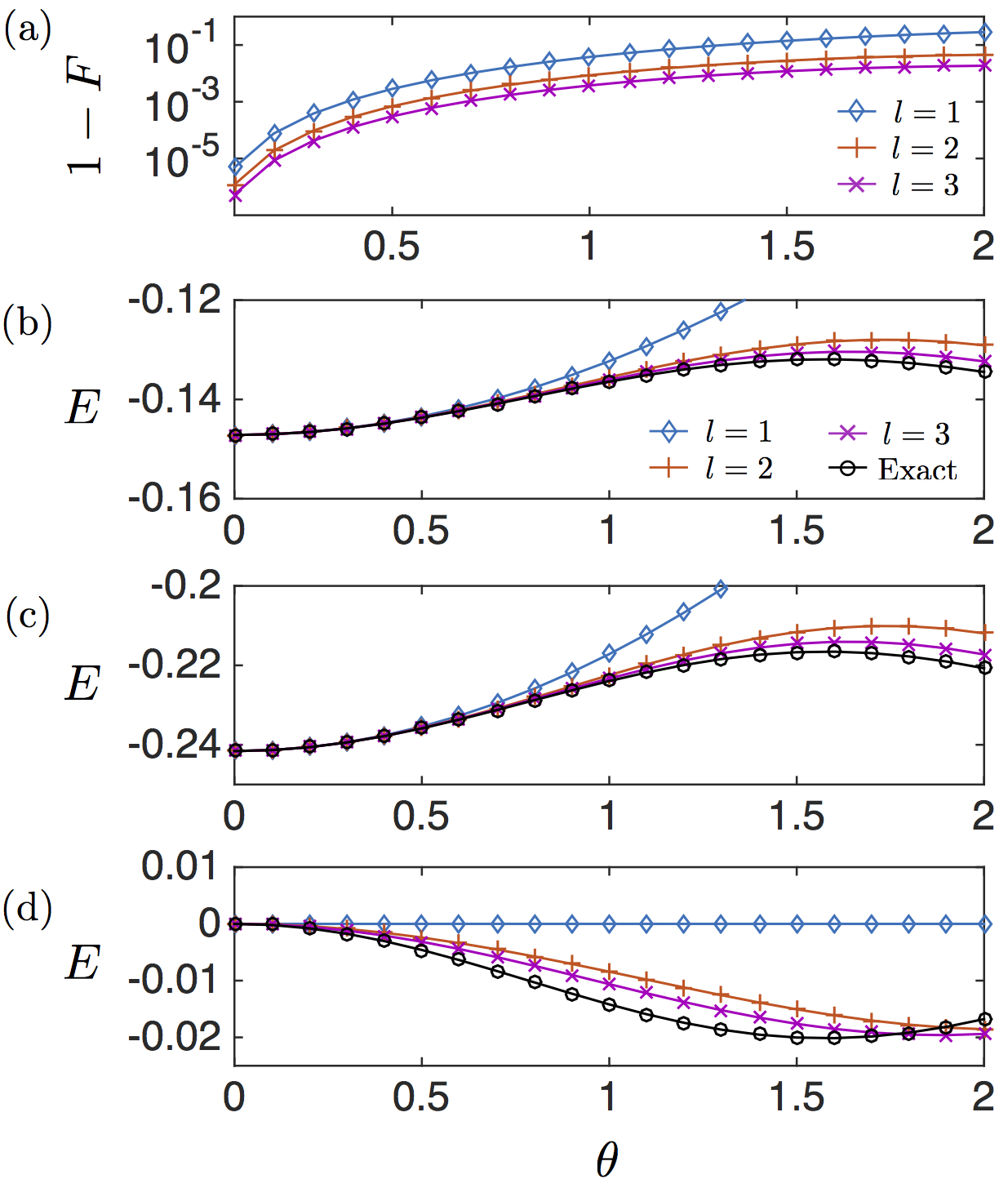}
\caption{Digital quantum simulation of the $\rm{H}_2$ molecule Hamiltonian for a phase of $\theta=h_{11}t$. Here, digital evolutions up to 3 Trotter steps are compared with the exact evolution for initial state $|\psi\rangle = c^{\dag}_1 c^{\dag}_2 |{\rm{vac}}\rangle = |1 1 0 0\rangle$. (a) Fidelity loss of the digitally evolved states, with $F=|\langle\Psi(t)|\Psi_l(t)\rangle|^2$. Expectation values of the separated Hamiltonians, in atomic units, proportional to (b) $\sigma_1^z$ and $\sigma_2^z$, (c) $\sigma_3^z$ and $\sigma_4^z$, and (d) $\sigma_1^x\sigma_2^y\sigma_3^y\sigma_4^x$, $\sigma_1^y\sigma_2^x\sigma_3^x\sigma_4^y$, $\sigma_1^x\sigma_2^x\sigma_3^y\sigma_4^y$ and $\sigma_1^y\sigma_2^y\sigma_3^x\sigma_4^x$.}
\label{Plots}
\end{figure}

In our case, we consider the evolution operators associated with the different summands of the Hamiltonian in equation~(\ref{HamSpin}), which corresponds to the sequence of gates in Fig.~\ref{Tstep}. We propose an algorithm based on the optimised tunable ${\rm{CZ}}_\phi$ gate, which allows one to perform efficiently ZZ interactions, or XX interactions in our basis~\cite{Barends15}. In this sense, we arrange the gates and the simulated interactions such that it allows us to simplify the algorithm by eliminating some entangling gates and their inverses, as shown in Fig.~\ref{Tstep}. The single Trotter step depicted in this figure represents the approximated evolution for a time $t/l$ of the complete Hamiltonian. Note that the third and fourth logical qubits correspond to the fourth and third physical qubits, respectively. We choose this notation due to the reduction of SWAP gates needed for the performance of the protocol. The optimized Trotter step contains 24~${\rm XX}$ two-qubit gates between nearest-neighbour qubits, 24~${\rm SWAP}$ gates and 20~single-qubit rotations. In Fig.~\ref{Plots}, we show the efficiency of the digital protocol for different number of Trotter steps. Here, we analyze the loss of the state fidelity and the expected value of some operators performed in the simulation, considering simulated phases up to $\theta=h_{11}t=2$. We break down the Hamiltonian terms and plot the energies of each of them to observe separately the Trotter error associated with the different kinds of interactions appearing in the algorithm. We observe that, for a single Trotter step, the energies related to single-qubit gates are similar to the exact evolution, while in the case of the four body terms the deviation is higher.

Symmetric Trotter expansions provide the improvement of the digital error at the expense of more gate execution. The Hamiltonian of equation~(\ref{HamSpin}) can be divided in two groups of interactions, $H_1$, the sum of the first 10 terms that commute among them, and $H_2$, the sum of the last 4 terms that also commute among them. As commuting interactions do not generate digital error, the evolution of a symmetric Trotter step can be written as follows~\cite{Suzuki90},
\begin{equation}
e^{-iH_1t/2l}e^{-iH_2t/l}e^{-iH_1t/2l}.
\end{equation}

\begin{figure}
\centering
\includegraphics[width=9.5cm]{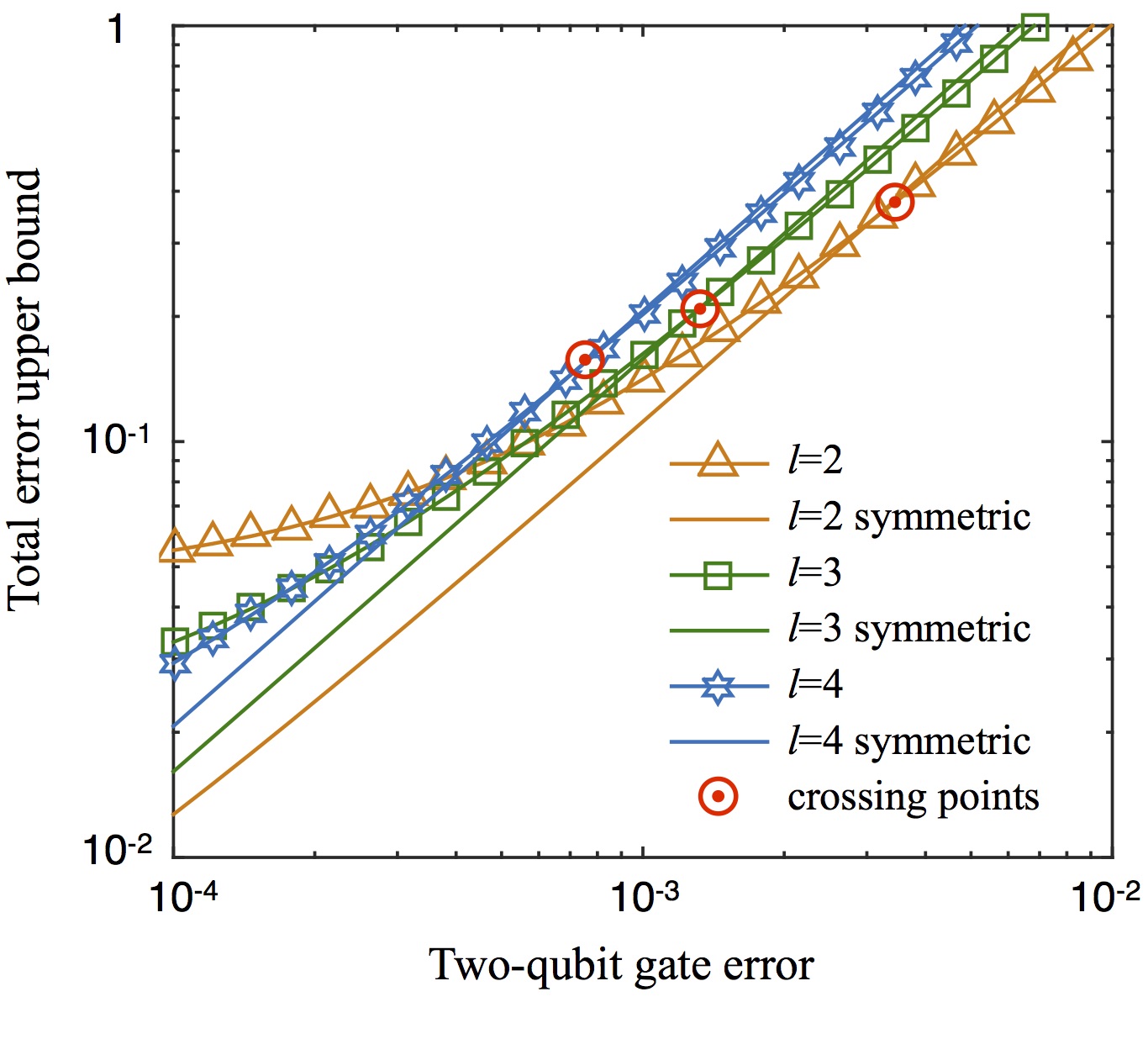}
\caption{Total upper bound of symmetric and regular expansions for the digital simulation of the hydrogen molecule as a function of the error of a two-qubit gate, considering $l=2,3,4$ Trotter steps and a simulated phase of $\theta=h_{11}t=2$. The total error is calculated as the sum of the experimental error of two-qubit gates and the digital errors. This plot shows the crossing points between the symmetric and the regular protocols for the same number of Trotter steps. On the left hand side of the crossing points, the symmetric protocol provides better results than the regular one. On the right hand side, however, the considered experimental gate error is higher, and the regular protocol where a less number of gates is executed shows better performance.}
\label{TrotterSym}
\end{figure}

This protocol requires the implementation of the interactions of $H_1$ one more time than in the regular digital protocol, thereby reducing the digital error. We introduce a fixed error for any two-qubit gate between nearest-neighbour qubits, without restricting ourselves to a specific setup or experimental source of error. Single-qubit gate errors are neglected due to their high fidelity with current technology. If the dominating error is the experimental one, then the aim is to reduce the number of gates and, consequently, the regular protocol gets better fidelities. In Fig.~\ref{TrotterSym}, we analyze the errors of both the regular Trotter protocol and the symmetric protocol, and we give an upper bound of the total error summing the digital and the experimental error considering a range of values for the two-qubit gate error employed. For fixed number of Trotter steps, $l=2,3,4$, we observe crossing points between the errors associated with the symmetric protocol and the regular one whilst considering higher experimental gate error. On the left side of the crossing points, the experimental error is smaller and the symmetric protocol provides better results, whereas on the right side, as the experimental gate error grows, the regular protocol is more adequate. We also notice that, as the number of Trotter steps increases, the advantages of one protocol with respect to the other lessen. It is worthy to mention that the two-qubit gate errors are on the order of $10^{-2}$ in superconducting devices~\cite{Barends2014}.

\subsection*{Simulation of environmental effects}
In this section, we propose a quantum simulation in superconducting circuits of generic system-environment interactions, which have long been recognised as fundamental in the description of electron transport in biomolecules. 

Biological systems are not isolated, and one can consider minimal models for characterising the quantum baths and decoherence~\cite{QAspectsLife,ChargeEnergyTrans}, such as the spin-boson model, or the Caldeira-Leggett model. The former is a widely used model that describes the interaction between a two-level system and a bosonic bath, and the latter deals with the dynamics of a quantum particle coupled to a bosonic bath. Usually, the coupling of the quantum system to the bath degrees of freedom is completely specified by the spectral density $J(\omega)$, which may be obtained from experimental data, and allows us to explore different continuum models of the environment. Nevertheless, in certain limits of strong coupling, the evaluations are computationally hard, and the complete comprehension of the physics remains as an open problem.

In particular, we study a Hamiltonian describing the charge transfer in DNA wires~\cite{Bulla06,Tornow06,Cuniberti09}, where experiments show a wide range of results, from insulator to conductor behaviours~\cite{Porath00,Storm01,Yoo01,Schuster04,Xu04,Cohen05,Nogues06}. When describing the dynamics of electrons in these biomolecules, the influence of a dissipative medium determines substantially the transfer events. We consider a bosonic bath in which a variety of crucial factors are contained, such as the internal vibrations of the biomolecule and the environmental effects.

\begin{figure}[t!!!]
\centering
\includegraphics[width=9.5cm]{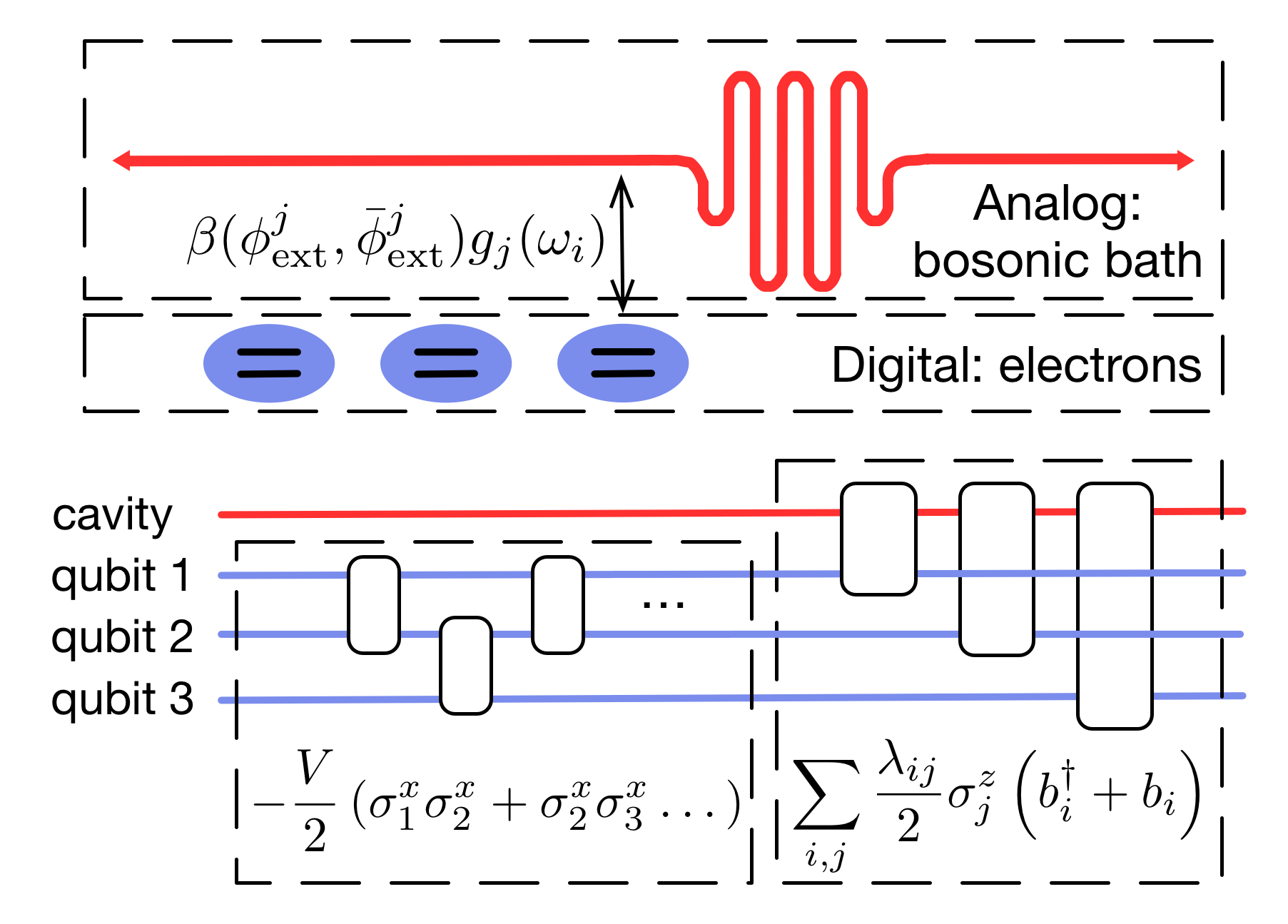}
\caption{Scheme of the cQED setup and digital-analog protocol needed for mimicking the Hamiltonian in equation~(\ref{spin-bath}). We consider a multimode cavity (red), that is, either a long resonator, a 3D cavity or a transmission line, coupled to three tunable superconducting qubits (blue). The cavity simulates analogically the bosonic bath, whereas the electrons are encoded in the superconducting qubits. The coupling between the qubits and the cavity, $\beta(\phi^j_{\rm ext},\bar{\phi}^j_{\rm ext}) g_j(\omega_i)$, must be tunable via external magnetic fluxes $\phi^j_{\rm ext}$ and $\bar{\phi}^j_{\rm ext}$ to enable the digital-analog quantum simulation, in which each qubit is coupled sequentially with the cavity.}
\label{CircuitScheme}
\end{figure}

A generic effective charge-bath model that describes an electronic system coupled to a fluctuating environment, in this case a bosonic bath, is captured by the Hamiltonian~\cite{Cuniberti09}
\begin{eqnarray} \label{charge-bath}
H = && \sum_j \varepsilon_j f^{\dag}_j f_j + \sum_j V_{j,j+1}\left(f^{\dag}_j f_{j+1} + \text{H.c.} \right) \nonumber \\
&& + \sum_i \omega_i b^{\dag}_i b_i + \sum_{i,j} \lambda_{ij} f^{\dag}_j f_j \left(b^{\dag}_i + b_i \right),
\end{eqnarray}
with $f_j(f^{\dag}_j)$, fermionic annihilation (creation) operators for electrons on different sites with energies $\varepsilon_j$. $V_{j,j+1}$ characterises the electron hopping between nearest-neighbour sites. The bath is represented by the bosonic annihilation (creation) operators $b_i(b^{\dag}_i)$, and the coefficients $\lambda_{ij}$ indicate how the system and bath are coupled.

A minimal and particular case is the two-site model with $j=A,B$, which comprises a donor (D) and an acceptor (A) site~\cite{Bulla06,Tornow06}. This reduced model can be mapped onto the spin-boson model, which has been studied in cQED~\cite{Haeberlein15}, for the particular case of one spinless electron in the system. We provide the patterns to treat in a cQED setup a more general situation where the spin degree-of-freedom or more electrons enter into the description. To this end, we consider equation~(\ref{charge-bath}) with $j=1,2,3$, and $V_{j,j+1}=V$, which cannot be mapped onto the well-studied spin-boson model. For the sake of simplicity, we have chosen this truncation, but the techniques can be easily extrapolated to an arbitrary case. 

As previously shown, in order to simulate fermionic operators in superconducting circuits, we replace them by Pauli matrices via the Jordan-Wigner transformation, leading to
\begin{eqnarray}\label{spin-bath}
H = && \frac{1}{2} \sum^3_{j=1} \varepsilon_j \sigma^{z}_j - \frac{V}{2} \big(\sigma^{x}_1 \sigma^{x}_2 + \sigma^{y}_1 \sigma^{y}_2 + \sigma^{x}_2 \sigma^{x}_3 + \sigma^{y}_2 \sigma^{y}_3 \big) \nonumber \\
&& + \sum_i \omega_i b^{\dag}_i b_i + \sum_{i,j} \frac{\lambda_{ij}}{2} (\sigma^{z}_j + 1) \left(b^{\dag}_i + b_i \right),
\end{eqnarray}
where the first two terms correspond to the purely electronic subsystem, the third term is the free energy of the bosons in the bath, and the last term represents the interaction of the electrons with the environment.

The Hamiltonian is now suitable for a digital quantum simulation in superconducting circuits, in which the qubits are described by Pauli operators, and 3D cavities, multimode coplanar waveguides, or low-Q cavities play the role of bosonic baths. A first step in this direction, considering an open transmission line coupled to qubits in order to simulate fermionic systems interacting with a continuum of bosons was introduced in the context of quantum field theory~\cite{Laura15}. While the basic protocol was already developed in this article, here we apply this formalism to the different context of electron transport in biomolecules, for a discrete set of coupled fermionic and bosonic modes. Recently, experimental realisations with a transmon qubit coupled to a multimode cavity in the strong coupling regime have been performed~\cite{Sundaresan15}. There, the feasibility of coupling a superconducting transmon qubit to a long coplanar resonator has been shown, achieving in this way the coupling of a qubit to a set of several bosonic modes at the same time. This multimode treatment is also needed to explain results in superconducting 3D cavities or in transmission lines~\cite{Nigg12,McKay15}, which allows us to propose a simulation exploiting the natural complexity that superconducting circuits reveal.

By coupling three tunable superconducting qubits~\cite{Srinivasan11,Chen14} to a multimode cavity as in Fig.~\ref{CircuitScheme}, the Hamiltonian of equation~(\ref{spin-bath}) can be reproduced by using digital-analog methods, that is, introducing the fermionic interactions digitally and the bosonic ones in analog interaction blocks. We propose the emulation of a variety of system-environment dynamics on superconducting circuit technology. To this end, we consider the interaction term describing the $j$th qubit coupled to a multimode cavity, 
\begin{equation}
H_{\rm int} = \sum_i \beta\left(\phi^j_{\rm ext},\bar{\phi}^j_{\rm ext} \right) g_j(\omega_i)\sigma^z_j (b_i^\dagger +b_i), 
\end{equation}
with $b_i$ $(b_i^\dagger)$ the $i$th mode annihilation (creation) operator related with the cavity frequency $\omega_i$, couplings $g_j(\omega_i) = g_0 \sqrt{i+1}$, and $g_0$ the coupling strength to the fundamental cavity mode $\omega_0$. We profit from the tunability of the coupling between qubits and transmission lines via external magnetic fluxes $\phi^j_{\rm ext}$ and $\bar\phi^j_{\rm ext}$~\cite{Laura15,Srinivasan11,Chen14} to address a wider range of regimes and models, since the set of couplings $\beta(\phi^j_{\rm ext},\bar{\phi}^j_{\rm ext} ) g_j(\omega_i)$ mimic the coefficients $\lambda_{ij}$ that characterise the interaction with the environment in equation~(\ref{spin-bath}). Moreover, it has been shown experimentally how to engineer different shapes for the bath spectral function with a transmission line and partial reflectors~\cite{Haeberlein15,FornDiaz16}. Additionally, it can also be proven that a simple tunable Ohmic bath, as the one provided by a transmission line, equipped with a feedback protocol, can produce highly non-Markovian dynamics~\cite{Paul15}. Growing in electronic complexity in equation~(\ref{charge-bath}) implies adding more qubits coupled to the transmission line in Fig.~\ref{CircuitScheme}. However, we can take full advantage of the same multimode cavity by encoding the bath in a similar fashion. Hence, the cQED setup may be easily scaled up by coupling more qubits to the same transmission line.

Let us discuss how the Hamiltonian in equation~(\ref{spin-bath}) is decomposed into different digital and digital-analog blocks for the quantum simulation. As in the previous subsection, the purely electronic subsystem can be decomposed in Trotter steps and reproduced by single- and two-qubit gates. Since the bosonic operators do not enter in this part, we must decouple the tunable qubits from the transmission line to perform the required gates. The remaining terms are encoded in digital-analog blocks, where we divide the dynamics in different Trotter steps in which the multimode cavity enters in an analog way, providing the free energy of the bosons, and simulating the last term of equation~(\ref{spin-bath}). This last term is composed of purely bosonic interactions proportional to $(b_i^{\dag}+b_i)$, which may be simulated through a microwave driving in the cavity. It also involves qubit-boson interactions, $\sigma_j^z(b_i^{\dag}+b_i)$, which emerge from the coupling of each qubit with the multimode cavity, as in Ref.~\cite{Laura15}. A future analysis of the error in this protocol may include not only the error of the two-qubit gates, but also the contribution of the imperfect decoupling of the qubits with the transmission line in the single-qubit gates in which we introduce the coupling with the bosonic degrees of freedom. The figure of merit in the simulation is the final state of the fermions, which may be extracted from the final state of the qubits. The quantum simulation may be performed for different final times, thereby allowing us to reconstruct the electronic dynamics, such as transfer of excitations.

With this proposal, we have provided not only a way of extracting results illustrating different charge transport regimes in biomolecules, but also a way of testing different minimal models for describing molecules embedded in a bosonic environment. Superconducting circuits are a controllable quantum platform in which we can tune couplings between spins and bosons, and manipulate external conditions to engineer different baths. We analyse models of biological systems with a certain complexity and translate them to a controllable superconducting device that enjoys a similar complexity.

\section*{Discussion}
We have proposed methods to perform feasible digital and digital-analog quantum simulations of molecular structures and biomolecules with the state-of-the-art of superconducting circuit technology. We analyse different quantum chemistry models by increasing gradually the complexity, moving from purely fermionic models of molecular structures to descriptions of charge transport in biomolecules embedded in a bosonic medium. We aim to profit from the unique features of cQED, such as the strong coupling of a two-level system to bosonic modes, in order to represent controllable scenarios in which quantum chemistry and quantum biology models may be studied. The proposal includes a purely digital quantum simulation protocol for fermionic models, for which we provide general methods of encoding and the sequence of gates needed for the particular case of simulation of the ${\rm H}_2$ molecule. The previous formalism is partially used for simulating biomolecules affected by their bosonic surroundings, where we also add analog blocks with a multimode cavity playing the role of the bosonic bath, hence boosting the efficiency of quantum algorithms for quantum chemistry.

\section*{Acknowledgements}
We acknowledge support from two UPV/EHU PhD grants, UPV/EHU Project EHUA14/04, Basque Government IT472-10; Spanish MINECO FIS2012-36673-C03-02 and FIS2015-69983-P; Ram\'on y Cajal Grant RYC-2012-11391; and SCALEQIT EU projects.

\section*{Author Contribution}
L.G.-\'A. and U.L.H. designed the protocol in Figure 1, did the calculations and numerical analysis, and prepared the figures. L.G.-\'A., U.L.H., A.M., M.S, E.S, and L.L contributed developing ideas, analysing results and writing the manuscript.

\section*{Additional information}
The authors declare no competing financial interests.

\end{document}